

\magnification=1200 \parskip=7pt plus 1pt

\def\pd{\partial} \def\vr{\vert} \def\lra{\longleftrightarrow}
\def\ra{\longrightarrow}  \def\cf{{\cal F}} \def\Ra{\Rightarrow}

  \def\b{\beta} \def\d{\delta}
  \def\g{\gamma}
\def\G{\Gamma} \def\l{\lambda}  \def\x{\xi}
\def\y{\eta}

\def\mfock{{\cal F}^M}  \def\lfock{{\cal F}^L}
\def\mcharge{\xi^M} \def\lcharge{\xi^L} \def\mmoment{\eta^M}
\def\lmoment{\eta^L} \def\gener#1#2{{\cal O}_{#1,#2}}
\def\matter{\phi^M} \def\liouv{\phi^L}

\def\bbz{Z\!\!\!Z} \def\bbn{I\!\!N} 
\def\bbc{C\kern-6.5pt I}

\rightline{IC/92/17 ~~(January 1992)}

\bigskip

\centerline{{\bf SO(2,C) ~INVARIANT RING STRUCTURE OF BRST
COHOMOLOGY}}

\centerline{{\bf AND SINGULAR VECTORS IN 2D GRAVITY WITH ~C $<$
1~ MATTER}}

\bigskip\bigskip\bigskip\bigskip

\centerline{N. Chair, ~~V.K. Dobrev$^{*}$\footnote{}{$^{*}$
permanent address~: Bulgarian Academy of Sciences, Institute of
Nuclear  Research and Nuclear Energy, 72 Boul. Trakia, 1784
Sofia, Bulgaria.} ~and ~~H. Kanno}

\bigskip

\centerline{International Centre for Theoretical Physics}
\centerline{P.O.Box 586, 34100 Trieste, Italy}
\bigskip\bigskip

\centerline{ABSTRACT}

We consider BRST quantized 2D gravity coupled to conformal matter
with arbitrary central charge $c^M = c(p,q) < 1$ in the conformal
gauge. We apply a Lian-Zuckerman $SO(2,\bbc)$ ($(p,q)$ -
dependent) rotation to Witten's $c^M = 1$ chiral ground ring. We
show that the ring structure generated by the (relative BRST
cohomology) discrete states in the (matter $\otimes$ Liouville
$\otimes$ ghosts) Fock module may be obtained by this rotation.
We give also explicit formulae for the discrete states.  For some
of them we use new formulae for $c <1$ Fock modules singular
vectors which we present in terms of Schur polynomials
generalizing the $c=1$ expressions of Goldstone, while the rest
of the discrete states we obtain by finding the proper
$SO(2,\bbc)$ rotation. Our formulae give the extra physical
states (arising from the relative BRST cohomology) on the
boundaries of the $p \times q$ rectangles of the conformal
lattice and thus all such states in $(1,q)$ or $(p,1)$ models.

\bigskip\bigskip

\baselineskip=14pt

{\bf Introduction.}

Recently the continuum approach to 2D gravity [1-4] has received
a lot of attention. In particular, there was progress in the
computation of the BRST cohomological extra discrete states of 2D
gravity coupled to $c^M \leq 1$ conformal matter in the conformal
gauge [5-14]. (For the calculation of correlators and fusion
rules see [15].) More recently Witten [12] made the observation
that for $c^M =1$ the discrete states of ghost number $(-1)$,
parametrized by two negative integers $(r,s)$, give rise to a
polynomial ring of ghost number zero operators. This ring is
generated by two elements which correspond to the two cohomology
states with $(r,s) =$ (-1,-2), (-2,-1). The states of ghost
number zero give rise to spin one currents which act as
derivatives of this ring and obey the $W_\infty$ - algebra of
area preserving diffeomorphisms.

Not much is known about this ring for $c^M <1$. (Some conjectures
are made in [16] in the case of minimal conformal matter.) This
is the first question we address in this letter.  We use the
$SO(2,\bbc)$ rotation trick [6,17,10] to change $c^M =1$. The
operators of BRST cohomologies for $c^M < 1$ are obtained by
applying the $SO(2,\bbc)$ to those for $c^M = 1$. We observe that
the (chiral) ring structure, defined by the short distance
behaviour of operator product, is preserved by this rotation.
This includes also the spin one currents which again fall in
$SU(2)$ multiplets and generate a $W_\infty$ - algebra. This is
done in Section 1.

The second question we address is the construction of explicit
representatives of the relative BRST cohomology for $c^M <1$.  In
the case $c^M =1$ prominent role was played [7,8,13,11,14] by the
singular vectors of Fock spaces [18].  We try to generalize these
formulae, in particular, for the ghost number zero states. We
generalize the Goldstone formula for Fock singular vectors for
the cases $(r,s) = (1,s), (2,s), (r,1), (r,2), (3,3)$ and
arbitrary central charge, and we give a conjecture for the
general case. The singular vectors are given as linear
combinations of Schur polynomials $S_{k_1 , \ldots, k_n}$, ($ n =
r$ or $n=s$), of all (in general) ordered partitions $k_1 \geq
\ldots \geq k_n \geq 0$, $k_1 + \ldots + k_n = rs$, in contrast
to the $c=1$ case in which $k_1 = \cdots = k_r = s$, or $r \lra
s$. Our formulae for the discrete states have direct application
for the {\bf (1,q) models}, (or for the $(p,1)$ models). In the
general case our formulae are valid for (primary) fields on the
boundaries of the Kac table.  All this is presented in Section
2., while compact formulae for the action of Virasoro generators
on the Schur polynomials (needed for the derivation of singular
vectors) are derived in the Appendix.

\bigskip

{\bf 1. ~~SO(2,C) invariance of the ring structure of the
discrete states}

{\bf 1.1.} ~~ We can consider the BRST cohomology of the Virasoro
algebra $H^* ( Vir, {\cal M})$ for any Virasoro module $\cal M$
with the energy momentum tensor $T(z)$ of the central charge
$c=26$. The BRST operator $$ d ~=~ \int {dz\over 2\pi i} : (T(z)
+ {1\over 2} T^G (z) )c(z): \eqno{(1.1)} $$ acts on ${\cal
M}\otimes{\cal F}^G$, where ${\cal F}^G$ is a fermion Fock module
of $(b,c)$ ghosts with the energy momentum tensor $T^G(z)$. We
will be concerned with the case in which $\cal M$ is a tensor
product of two Feigin-Fuchs modules, ${\cal M}~=~\mfock \otimes
\lfock$.  In the free field realization of 2D quantum gravity
coupled to conformal matter, $\mfock$ corresponds to the matter
and $\lfock$ represents the Liouville field. We will distinguish
between the quantities for matter and the Liouville fields by
using the superscript $M$ and $L$, respectively.

Let us recall that the Feigin-Fuchs module $\cal F$ is a boson
Fock module of 2D scalar field $\phi(z)$. $\cal F$ is
characterized by the two parameters: the background charge $\xi$
and the momentum $p$ of the vacuum.  The energy momentum tensor
$$ T(z) ~=~ -{1\over 2} :\partial\phi(z)\partial\phi(z): ~+~ i\xi
\partial^2\phi((z) \eqno{(1.2)} $$ defines an action of Virasoro
algebra.  The central charge $c$ and the conformal weight $h$ of
the vacuum are given by $$ c ~=~ 1 - 12 \xi^2~, \qquad h ~=~
{1\over 2} p(p - 2\xi) ~=~ {1\over 2} (\eta + \xi)(\eta -
\xi) ~. \eqno{(1.3)} $$ Here we have introduced the "shifted"
momentum $\eta ~=~ p - \xi$, which will be used in the following
instead of the "unshifted" momentum $p$. We will denote the
Feigin-Fuchs module with the parameters $\xi$ and $\eta$, ${\cal
F}~(\xi,\eta)$.

The BRST cohomology of the Virasoro algebra for ${\cal
M}~=~\mfock (\mcharge,\mmoment) \otimes \lfock
(\lcharge,\lmoment)$ was studied in [5,6,11,14]. Let us fix the
matter central charge $c^M~=~1 - 12(\mcharge)^2$. The condition
$c^M + c^L =  26$ gives $$ \xi_{+}\xi_{-} ~=~ -1~, \eqno{(1.4)}
$$ where we have defined the light-cone combination
$$\eqalign{\xi_{\pm} ~=&~ {1\over {\sqrt 2}}(\mcharge \pm i
\lcharge)~, \cr  \eta^{\pm} ~=&~ {1\over {\sqrt 2}}(\mmoment \pm
i \lmoment)~. \cr} \eqno{(1.5)} $$ $\mcharge$ and $\lcharge$ are
fixed up to sign. In terms of two (real) parameters $r$ and $s$
defined by $$ r ~=~ -{1\over 2}\xi_{-}\eta^{+} ~, \qquad s ~=~
-{1\over 2}\xi_{+}\eta^{-} ~. \eqno{(1.6)} $$ one can parametrize
the momenta in the following way: $$ \eqalign{ \mmoment(r,s)~ =&~
{1\over 2} \Big[ (r+s)\mcharge + (r-s)i\lcharge \Big]~, \cr
i\lmoment(r,s)~ =&~ {1\over 2} \Big[ (r-s)\mcharge +
(r+s)i\lcharge \Big]~. \cr} \eqno{(1.7)} $$ Then the result of
[11] (for $c^M =1$ see [6]) is summarized as follows:

{\bf Theorem.} ~The relative cohomology $H^*_{rel}~(Vir,{\cal
M})$ is non-trivial only in the following cases. \item{(i)} If
$rs=0$, then $H^*_{rel} ~=~ \delta_{n,0}\bbc$~. \item{(ii)} If
both $r$ and $s$ are positive integers, then $H^*_{rel} ~=~
\delta_{n,0} \bbc \oplus\delta_{n,1} \bbc$~. \item{(iii)} If both
$r$ and $s$ are negative integers, then $H^*_{rel} ~=~
\delta_{n,0} \bbc \oplus \delta_{n,-1} \bbc$~.

\noindent In all cases the non-trivial cohomology states appear
at level $rs$.

The gravitationally dressed primary states of [3,4] are included
in the case (i). The cases (ii) and (iii) give the "extra"
physical states. (In fact there is a doubling of the physical
states since finally one has to take the absolute cohomology into
account (cf. [11]).) Especially, when $c^M = 1~(\mcharge = 0)$,
the physical states with ghost number $(-1)$ are basic
ingredients for the construction of the chiral ground ring by
Witten [12].  At the (discrete) momenta $\mmoment ~=~ \pm
(r-s)/\sqrt{2}$~, ~$\lmoment ~=~ \mp i(r+s)/\sqrt{2}$ with $r$
and $s$ being negative integers, there is an operator
$\gener{r}{s}$ corresponding to the BRST cohomology class of
ghost number $(-1)$.  $\gener{r}{s}$ has conformal spin zero and
ghost number zero, due to a shift of quantum numbers
corresponding to the ghost $c(z)$ in the transition from states
to operators. By this crucial property the short distance
behavior of the product $\gener{r}{s} (z)\gener{r'}{s'} (w)$
defines a ring structure of operators $\gener{r}{s}$.  Witten has
proved that this ring is the polynomial algebra generated by
$x=\gener{-2}{-1}$ and $y=\gener{-1}{-2}$ and that we can
identify $\gener{r}{s}$ with $x^{-r-1}~y^{-s-1}$. Hence the ring
structure is given by $$ \gener{r}{s} \gener{r'}{s'} ~=~
\gener{r+r'+1}{s+s'+1}~. \eqno{(1.8)} $$ (Note that the indeces
are negative integers.) Furthermore, at the same momenta as
$\gener{r}{s}$ we have a current $W_{r,s}(z)$ with ghost number
zero. $W_{r,s}(z)$ is also an extra BRST cohomology class (see
the case (iii) above.). $W_{r,s}(z)$ acts on the ground ring as a
derivation. The action is defined again in terms of the operator
product expansion (OPE): $$ W_{r,s}(z)\gener{r'}{s'} (w) ~\sim
{}~{\gener{r+r'+1}{s+s'+1} (w) \over z-w} ~+~ \cdots ~~,
\eqno{(1.9)} $$ where dots represent BRST exact terms.  From the
viewpoint of $SU(2)$ symmetry it is natural to include the
tachyonic currents $W_{0,s}$ and $W_{r,0}$. The relation (1.9) is
also valid for $r=0$ or $s=0$. Finally the symmetry algebra of
the ground ring is identified with $W_\infty$ - algebra of area
preserving diffeomorphisms [19]: $$ W_{r,s}(z) W_{r',s'}(w) ~\sim
{}~{W_{r+r'+1,s+s'+1}(w)\over z-w} ~. \eqno{(1.10)}$$

{\bf 1.2.} ~~ The existence of extra BRST cohomologies at
discrete values of momenta is independent of the value of
background charges for Feigin-Fuchs modules, as long as the
constraint (1.4) is satisfied.  Therefore, one can obtain the
operators $\gener{r}{s}$ and the currents $W_{r,s}$ for any value
of $c^M$ in the same manner as described above.  To investigate
the OPE relations of the types (1.8) $\sim$ (1.10), we will make
use of the $SO(2,\bbc )$ rotation of the form $$ R(m,n) ~=~
{1\over 2\sqrt{mn}} \left(\matrix{ m+n & i(m-n) \cr -i(m-n) & m+n
\cr}\right) ~. \eqno{(1.11)} $$ Any pair of background charges
$(\mcharge,\lcharge)$ satifying the condition (1.4) can be
obtained by making an appropriate $SO(2,\bbc )$ rotation from the
"reference" charges $(0,\pm i\sqrt2)$ for $c^M=1$ case. For
example, the background charges corresponding to $(p,q)$ model
are $$ \vec{\xi} ~=~ \left(\matrix{\mcharge \cr \lcharge
\cr}\right) ~=~ R(q,p) ~\left(\matrix{0\cr \pm i\sqrt2
\cr}\right) ~=~ \left(\matrix{ {\pm (p-q)\over \sqrt{2pq}} \cr
{\pm i(p+q)\over \sqrt{2pq}} \cr}\right) ~. \eqno(1.12) $$ In the
following we will use a vector notation like
$\vec{\xi}~=~(\mcharge, \lcharge)^t$ for a pair of quantities for
matter and the Liouville systems. The discrete momenta where we
have extra physical states are related to the background charges
by (1.7) with {\it integers} $r$ and $s$. It is quite remarkable
that this relation is decomposed into the product of rotation and
scaling: $\vec{\eta}(r,s) ~=~ \sqrt{rs}~R(r,s)~\vec{\xi} ~,$
which is valid for any value of $\vec{\xi}$. The discrete momenta
$\vec{\eta} (r,s)$ are obtained from those for $c^M=1$ case (the
"reference" momenta) by the {\it same} rotation as (1.12):
$$\vec{\eta} (r,s)  ~=~ \sqrt{rs} R(r,s)R(q,p)\left( \matrix{0
\cr  \pm i\sqrt{2}\cr}\right) ~=~ R(q,p)~\left( \matrix{ \pm
{1\over \sqrt{2}} (r-s) \cr \pm{1\over i\sqrt{2}} (r+s)
\cr}\right) ~~.  \eqno{(1.13)} $$ ($\vec{p} ~=~ \vec{\eta} +
\vec{\xi}$ is also rotated by the same  $R(q,p)$.)

The operators $\gener{r}{s}$ and the currents $W_{r,s}$ arising
from BRST cohomology for ${\cal M}=\mfock\otimes\lfock$ have
dependence on the characterestic parameters $\vec\xi$ and
$\vec\eta$ of $\cal M$, $$ \gener{r}{s} ~=~ \gener{r}{s}
(\vec{\xi},\vec{\eta}),\qquad W_{r,s} ~=~
W_{r,s}(\vec{\xi},\vec{\eta}) ~. \eqno{(1.14)} $$ This dependence
is universal in the sense that the way of constructing BRST
physical states is independent of $\vec\xi$ with $\vec\xi$ and
$\vec\eta$ treated as free parameters. Let us define an action of
$R(m,n)$ by $$ R(m,n) \cdot \gener{r}{s} (\vec{\xi},\vec{\eta})
{}~=~ \gener{r}{s} (R(m,n)\vec{\xi},R(m,n)\vec{\eta}) ~,
\eqno{(1.15)} $$ and the same for $W_{r.s}$. Then the
universality stated above implies $$ \gener{r}{s}^{(p,q)} ~=~
R(p,q)\cdot \gener{r}{s}^{(1,1)}~, \qquad W_{r,s}^{(p,q)} ~=~
R(p,q)\cdot W_{r,s}^{(1,1)} ~, \eqno{(1.16)} $$ where the
superscript $(p,q)$ specifies the model. We will show some
examples later on. In the calculation of OPE, two point function
for free scalar fields $\phi^I(z)~(I=M,L)$: $$ \langle
\phi^I(z)~\phi^J(w) \rangle ~=~ -\delta^{IJ} ln(z-w)
\eqno{(1.17)} $$ is used for contractions. Since two point
function (1.17) is invariant under $SO(2,\bbc )$ rotation,we can
conclude that the coefficient of OPE are independent of $(p,q)$,
which means, for example, $$\eqalign{ W_{r,s}^{(p,q)}(z)
\gener{r'}{s'}^{(p,q)}(w) ~=&~ R(p,q)\cdot W_{r,s}^{(1,1)}(z)
R(p,q)\cdot \gener{r'}{s'}^{(1,1)}(w) ~=\cr =&~ R(p,q)\cdot \big(
W_{r,s}^{(1,1)}(z) \gener{r'}{s'}^{(1,1)}(w) \big) ~\sim \cr \sim
&~ {1\over z-w} ~R(p,q)\cdot \gener{r+r'+1}{s+s'+1}^{(1,1)}(w) ~=
\cr = &~ {1\over z-w} ~\gener{r+r'+1}{s+s'+1}^{(p,q)}(w)~. \cr}
\eqno{(1.18)} $$ Thus, $SO(2,\bbc )$ rotation preserves the OPE
structure of BRST cohomology classes of types (1.8) $\sim$
(1.10). The whole algebraic structure appearing in the $c^M=1$
case remains true for any $(p,q)$. For example, the algebra of
$\gener{r}{s}^{(p,q)}$'s is identified with the polynomial ring
generated by $x^{(p,q)} ~=~ \gener{-2}{-1}^{(p,q)}$ and
$y^{(p,q)} ~=~ \gener{-1}{-2}^{(p,q)}$. The currents
$W_{r,s}^{(p,q)}$ acts on this polynomial ring as derivations
like $W_{0,-1}^{(p,q)} ~=~ \partial / \partial x^{(p,q)}$ and
$W_{-1,0}^{(p,q)} ~=~ \partial / \partial y^{(p,q)}$.

{\bf 1.3.} ~~ We show a few examples. The generators $x^{(p,q)}$
and $y^{(p,q)}$ are operators corresponding to the BRST
cohomology class with ghost number $-1$ at level 2: $$ \biggl[
b_{-2} - {\xi_{-}\over\sqrt 2}\big(\matter_{-1} +i\liouv_{-1}
\big) b_{-1} \biggr] \vert \vec{\eta}(-2,-1) \rangle~,
\eqno{(1.19.a)} $$ and $$ \biggl[ b_{-2} - {\xi_{+}\over\sqrt
2}\big(\matter_{-1} -i\liouv_{-1}  \big) b_{-1} \biggr] \vert
\vec{\eta}(-1,-2) \rangle~, \eqno{(1.19.b)} $$ respectively.
Noting a shift of momenta in going to operators, we obtain $$
\eqalign{ x^{(p,q)} ~=&~ \biggl[ cb - {\xi_{-} \over\sqrt
2}i\big(  \partial\matter + i\partial\liouv\big)\biggr]
\exp{-i{\xi_{+} \over \sqrt 2}(\matter -i\liouv)}~,\cr
y^{(p,q)}~=&~\biggl[ cb - {\xi_{+} \over\sqrt 2}i\big(
\partial\matter - i\partial\liouv\big)\biggr] \exp{-i{\xi_{-}
\over \sqrt 2}(\matter +i\liouv)}~.\cr} \eqno{(1.20)} $$ The
dependence on $(p,q)$ only appears in the parameters $\xi_{\pm}$
on which $SO(2,\bbc )$ acts as scale transformation.  Our next
example is $SO(2,\bbc )$ rotated $SU(2)$ current algebra.  In
terms of a combination $$ \Phi^{(p,q)}~=~{\xi_{+} \over\sqrt 2}
\big(  \partial\matter - i\partial\liouv\big)  - {\xi_{-}
\over\sqrt 2} \big(  \partial\matter + i\partial\liouv\big)
\eqno{(1.21)} $$ the generators of $SU(2)$ current algebra is
given by $$ \eqalign{ T_+ ~=& ~W_{0,-2}^{(p,q)} ~=~ \exp
{i\Phi^{(p,q)}}~, \cr T_3 ~=&~ W_{-1,-1}^{(p,q)} ~=~ i\partial
\Phi^{(p,q)}~, \cr T_- ~=&~ W_{-2,0}^{(p,q)} ~=~ \exp
{-i\Phi^{(p,q)}}~. \cr} \eqno{(1.22)} $$ The $SU(2)$ currents
acts on the polynomial ring in $x^{(p,q)}$ and $y^{(p,q)}$ as
derivations: $$ \eqalign{ T_{+} ~=&~ x^{(p,q)}{\partial\over
y^{(p,q)}} ~, \cr T_3 ~=&~ {1\over 2}\big(
x^{(p,q)}{\partial\over\partial y^{(p,q)}} - y^{(p,q)}
{\partial\over\partial x^{(p,q)}}\big) ~, \cr T_{-} ~=&~
y^{(p,q)}{\partial\over\partial x^{(p,q)}}~. \cr} \eqno{(1.23)}
$$ The currents $W_{r,s}^{(p,q)}~~(r+s=-n)$ constitutes an
$SU(2)$ multiplet of spin $n\over 2$. The following is an example
of spin $3\over 2$ multiplet: $$ \eqalign{
W_{-3,0}^{(p,q)}~=&~\exp{{i\over\sqrt 2}\biggl( \xi_{-}(\matter +
i\liouv)-2(\xi_{+}(\matter -i\liouv)\biggr)} ~, \cr
W_{-2,-1}^{(p,q)}~=&~{1\over 2}\big(T_3^2 -\partial T_3\big)
\exp{-{i\over\sqrt 2} \xi_{+}(\matter-i\liouv)} ~, \cr
W_{-1,-2}^{(p,q)}~=&~{1\over 2}\big(T_3^2 +\partial T_3\big)
\exp{{-i\over\sqrt 2} \xi_{-}(\matter + i\liouv)} ~, \cr
W_{0,-3}^{(p,q)}~=&~\exp{{i\over\sqrt 2}\biggl( \xi_{+}(\matter -
i\liouv)-2(\xi_{-}(\matter +i\liouv)\biggr)} ~. \cr}
\eqno{(1.24)} $$ With these examples at lower levels it is easy
to check the OPE structure of type (1.8) $\sim$ (1.10)
explicitly.  The conditions necessary for this calculation are
the constraint (1.4) and the fundamental two point functions
(1.17).

\bigskip

{\bf 2. ~Discrete states in (p,q) models and Fock modules
singular vectors}

{\bf 2.1.} ~~~In this Section we give explicit representatives of
the discrete states of $H^*_{rel}~(Vir,{\cal M})$ for arbitrary
$(p,q)$ models. We start with the {\bf ghost number zero} case.
For $c^M = 1$ it is well known [7,8,13,11,14] that explicit
formulae for the physical states are provided by the singular
vectors of the matter Fock spaces expressed in terms of the so
called Schur polynomials which are recalled in the Appendix.

Let $\x^M$ be given as in (1.12) with $p,q \in \bbn$. Then the
Fock module $\cf^M(\x^M , \y^M)$ is reducible [20] iff $\y^M$ is
given by (1.7) with $r,s \in \bbz , rs >0$.

Let us consider first the case $r,s \in \bbn$. For $c^M = 1$ the
singular vector of level $rs$ is given by [18] (see also [11,14]
whose notation we partly follow): $$ v^{c=1}_{r,s} ~=~ S_{ {
\underbrace{s, \ldots, s}_r }} ~ \left( { \sqrt{2} \over n}
\y^M_{-n}\right) ~\vr \y^M \rangle ~, \eqno(2.1a)$$ or,
equivalently, by $$ v^{c=1}_{r,s} ~=~ S_{ \underbrace{r, \ldots,
r} \atop s} \left( - { \sqrt{2} \over n} \y^M_{-n}\right) ~\vr
\y^M \rangle ~, \eqno(2.1b)$$ the two expressions differing by
sign. We recall that a singular vector of a Virasoro highest
weight module (here Fock modules) is a weight vector ~$v$~
different from the highest weight vector and obeying $L_n v ~=~
0$, for $n \geq 1$, where $L_n$ are the Virasoro generators. Let
us note that the Schur polynomials involved are exactly of length
$r$, (resp.  $s$), i.e., each term involves the product of
exactly $r$, (resp.  $s$), elementary Schur polynomials.

We would like to generalize these formulae for $c < 1$. For $r=1$
or $s=1$ one may check that the singular vectors are given by $$
v_{1,s} ~=~ S_s \left({ \sqrt{2} \over n} \x_+ \y^M_{-n}\right)
{}~\vr \y^M \rangle ~, ~~ r=1 ~, \eqno(2.2a)$$ $$ v_{r,1} ~=~ S_r
\left( { \sqrt{2} \over n} \x_- \y^M_{-n}\right) ~\vr \y^M
\rangle ~, ~~ s=1 ~,  \eqno(2.2b)$$ which reduce to (2.1) for
$c=1$, $p=q=1$, $\x_\pm = \pm 1$. The derivation of (2.2) and the
formulae, necessary for the derivation of the singular vectors
given below, are given in the Appendix.

The cases $r,s >1$ are much more interesting. In particular, for
$r=2$ or $s=2$ we have: $$ v_{2,s} ~=~ \sum_{k=0}^s ~\b^s_k
\left( \x_+ \right) S_{ 2s-k , k } \left({ \sqrt{2} \over n} \x_+
\y^M_{-n}\right) ~\vr \y^M \rangle ~, ~~ r=2 ~, \eqno(2.3a)$$ $$
v_{r,2} ~=~ \sum_{k=0}^r ~\b^r_k \left( \x_- \right) S_{ 2r-k , k
} \left({ \sqrt{2} \over n} \x_- \y^M_{-n}\right) ~\vr \y^M
\rangle ~, ~~ s=2 ~, \eqno(2.3b)$$ $$ \b^n_k \left(\x_\pm \right)
{}~=~ { (-1)^k \G (2(n+u_\pm)) \over \G(2n -k + u_\pm +1 )
\G(k+u_\pm) } ~=~ (-1)^k \left({ 2(n+u_\pm) - 1 \atop k + u_\pm -
1 }\right) ~, \eqno(2.3c)$$ where ~$u_\pm ~=~ \g_\pm (\y^M -
\x^M) - 1$ ~$=~ \x_\pm^2 - n$, (cf. the Appendix), and only the
expression in terms of $\G$ - functions can be used if $u_\pm$ is
not integer.

Our first observation is that for $c < 1$ and $r=2$, (resp.
$s=2$), all Schur polynomials of degree $2n$ and of length $\leq
2$ are involved, since the term with $k=0$ involves the
elementary Schur polynomial $S_{2n} ~=~ S_{ 2n , 0 }$. Of course,
there are partial cases when not all terms are present in
(2.3a,b). Thus for $p=1$, resp. $q=1$, we have $u_+ = q -s$ for
(2.3a), resp. $u_- = p-r$ for (2.3b), and $\b^n_k (\x_\pm) ~=~ 0$
for $k < 1 - u_\pm $. Finally, for $c=1, p=q=1$ we have $\b^n_k
(\pm 1) ~=~ 0$ for $k < n$ and (2.3a,b) collapse (up to signs) to
(2.1a,b). Note that $\b^n_n ~\neq ~0$ in all cases.

Let us give also the example of a singular vector in the case
$r=s=3$ : $$v_{3,3} ~=~ \sum_{ 9-j-s \geq j \geq s \geq 0 }
{}~\b^9_{j,s} (\x_+) ~S_{9-j-s, j, s} \left( { \sqrt{2} \over n }
\x_+ \y^M_{-n} \right) ~\vr \y^M \rangle ~, \eqno(2.4a)$$
$$\b^9_{j,s} (\x_+) ~=~ { (-1)^{j+s} \b_{j,s} \over \G(10 - j - s
+ u ) \G(j+u) \G(s+u-1) u} ~, \eqno(2.4b)$$ $$\eqalign{ \b_{0,0}
{}~=~ \b_{1,0} ~=~ (u-3) (u+8) ~, ~~&\b_{1,1} ~=~ \b_{2,1} ~=~
(u-1) (u+6) ~, \cr ~~\b_{4,0} ~=~ \b_{4,1} ~=~ - (u+2) (u+3) ~,
{}~~&\b_{2,2} ~=~ \b_{3,2} ~=~ \b_{3,3} ~=~ u (u+5) ~, \cr
{}~~\b_{2,0} ~=~ -18 ~, ~~\b_{3,1} ~=~ -6 ~, ~~&\b_{3,0} ~=~ - (u^2
+ 5u +12) ~. \cr} \eqno(2.4c) $$ where ~$u ~=~ u_+ ~=~ \g_+
(\y^M(3,3) - \x^M) - 1$ ~$=~ 2\x_+^2 -3 ~=~ 2q/p - 3 ~>~ -3$,
(cf. the Appendix). Note that $\b^9_{j,s}$ have no poles (even
for $u=0$).  For $c^M =1$, $p=q =1 = -u$, all coefficients are
zero except $\b^9_{3,3}$ and (2.4) goes into (2.1). For $u= -2,
0, 1, 3$ there are also some vanishing coeffiecients (always
$\b^9_{0,0}$). Note that $\b^9_{3,3} \neq 0$ in all cases.  The
same expression is valid if we replace $\x_+ \ra \x_-$, $u = u_+
\ra u_- ~=$ ~$2\x_-^2 -3 ~=~ 2p/q - 3$.

Thus we are lead to the conjecture that the general expression
for the singular vectors should be: $$ v_{r,s} ~=~ \sum_{k_1 \geq
\ldots \geq k_r \geq 0 \atop k_1 + \ldots + k_r = rs} {
(-1)^{k_1} \b_{k_1 , \ldots , k_r} \left( \x_+ \right) \over \G
(k_1 + u_+ + 1 )  \ldots \G (k_r + u_+ - r + 2) } S_{k_1 , \ldots
, k_r} \left({ \sqrt{2} \over n} \x_+ \y^M_{-n}\right) \vr \y^M
\rangle \eqno(2.5a)$$  or, up to sign, $$ v_{r,s} ~=~ \sum_{k_1
\geq \ldots \geq k_s \geq 0 \atop k_1 + \ldots + k_s = rs} {
(-1)^{k_1} \b_{k_1 , \ldots , k_s} \left( \x_- \right) \over \G
(k_1 + u_- + 1) \ldots \G (k_r + u_- - s + 2) } S_{k_1 , \ldots ,
k_s} \left({ \sqrt{2} \over n} \x_- \y^M_{-n}\right) \vr \y^M
\rangle \eqno(2.5b)$$ where $u_\pm ~=~ \g_\pm (\y^M(r,s) - \x^M)
- 1$ ~$=~ \sqrt{2} \x_\pm (\y^M(r,s) - \x^M) - 1$, (cf.  the
Appendix), and the coefficients obey a simple recursion relation
(from the action of $L_1$): $$ \b_{k_1 , \ldots , k_n} ~=~
\sum_{j=2}^n ~\b_{k_1 -1 , \ldots , k_j +1, \dots, k_n} ~,
\eqno(2.5c)$$ and a more complicated one (from $L_2$) which we
omit for the lack of space. Note that in (2.5a), (resp. (2.5b)),
$s\leq k_1 \leq rs$, (resp.  $r\leq k_1 \leq rs$), $0\leq k_r
\leq s$, (resp. $0\leq k_s \leq r$), ~$\b_{ s, \ldots, s} \neq
0$, ~$\b_{ r, \ldots, r } \neq 0$. Thus according to our
conjecture Schur polynomials of degree $rs$, and of arbitrary
length $\leq r$, (or $\leq s$), are involved, i.e., every term
involves the product of at most $r$, (resp. $s$), elementary
Schur polynomials.

The {\bf Proof} ~that the states given in (2.2) - (2.5) are
representatives of $H^{(0)}_{rel} (Vir,{\cal M})$ is analogous to
the case $c^M = 1$ [11]. Since these are singular vectors they
are BRST closed. Consider (2.5b). It is enough to note that by
(2.2b), (2.3b), (2.4b), and our conjecture, the term $S_{ r,
\ldots, r} (\sqrt{2} \x_- \y^M_{-n}/n) ~\vr \y^M, \y^L \rangle$
is always present which is the representative for $c^M =1$ and
then we can repeat the proof of [11]. In the last step we are
reduced to the term $({\tilde \y}^+_{-r})^s = (\x_-)^s
(\y^+_{-r})^s$ which is representative of $H^{(0)}_{rel}
(Vir,{\cal M})$ [11].  Analogously is considered (2.5a).

Let us note that we could have used also the singular vectors of
the same degree in the Liouville sector.

Finally, let us remark that representatives for the case $c^M <1$
may be obtained from (2.1) by the $SO(2,\bbc)$ rotation of [6],
however, these would not be singular vectors.

{\bf 2.2.} ~~Next we consider the ghost number zero case with
{}~$r,s \in -\bbn$. The physical importance of these sates comes
from the fact that they are associated to the currents ~$W_{r,s}
{}~=~ W^{(p,q)}_{r,s}$~ discussed in Section 1.

First we note that for $c^M = 1$, $\x^M = 0$, from (1.6) we have
that $\y^M (r,s) =$ $\y^M (-s , -r)$. Thus the cohomology
representatives are singular vectors $v^{c=1}_{-s,-r}$ as given
in (2.1) (cf. also [14]).

For $c^M < 1$ things are rather more complicated with one
exception, namely the case $$r ~=~ -kp ~, ~~~s ~=~ -\ell q ~,
{}~~k,\ell \in \bbn . \eqno(2.6a)$$ In this case we have (cf.
(1.7)): $$\y^M (r,s) ~=~ \y^M (p\ell , kq) ~, \eqno(2.6b)$$ and
the cohomology representatives at level $r s = k \ell p q$ are
singular vectors $v_{p\ell , kq}$ from (2.5) (and also by (2.2) -
(2.4) whenever applicable). Let us remark that these $r,s$
correspond to corners of the conformal lattice with dimension
$p,q$ [21,22,23,20], (see also [24] for an explicit
parametrization of the reducible Virasoro highest weight
modules).

In the generic situation when (2.6a) is not fulfilled there is no
singular vector of degree $rs$ in the matter sector (and also in
the Liouville sector). (There is only the so called cosingular
vector [20].) Thus we can obtain representatives only by
implementing the proper $SO(2,\bbc)$ rotation of [6] as discussed
in Section 1. In this case we have the following representatives
$$\psi^{(0)}_{r,s} ~=~ S_{ { \underbrace{-r, \ldots, -r}_{-s} }}
(x) ~\vr \y^M, \y^L \rangle ~, ~~{\rm {\bf or}} \eqno(2.7a)$$
$$\psi^{(0)}_{r,s} ~=~ S_{ { \underbrace{-s, \ldots, -s}_{-r} }}
(-x) ~\vr \y^M, \y^L \rangle ~, \eqno(2.7b)$$ $$ x_n ~=~ { i
\over n} \left( \x^L \y^M_{-n} - \x^M \y^L_{-n} \right) ~.
\eqno(2.7c)$$ For $c^M = 1$, $\x^M = 0$, $\x^L = -i\sqrt{2}$,
formulae (2.7) go into (2.1).

{\bf 2.3.} ~~Finally we consider the cases of {\bf nonzero ghost
number}. In these cases the cohomology representatives are given
in terms of elementary Schur polynomials and the formulae may be
obtained from [14], Theorem 2.2., by the replacement ~$\y^\pm \ra
{\tilde \y}^\pm$~, for $r,s \in \pm \bbn$. We mention only the
partial cases $\vr r \vr = 1$ or $\vr s \vr =1$, since these are
simplified in comparison with [14]. We have: $$ \psi^{(1)}_{r,1}
{}~=~ \sum_{j=1}^r S_{ r-j} \left( - {\tilde \y}^+_{-n}/n \right)
c_{-j} ~\vr \y^M, \y^L \rangle ~, ~~r \in \bbn ~, \eqno(2.8a)$$
$$ \psi^{(1)}_{1,s} ~=~ \sum_{j=1}^s S_{s-j} \left( - {\tilde
\y}^-_{-n}/n \right) c_{-j} ~\vr \y^M, \y^L \rangle ~, ~~s \in
\bbn ~; \eqno(2.8b)$$ $$ \psi^{(-1)}_{r,-1} ~=~ \sum_{j=1}^{-r}
S_{- r-j}  \left( - {\tilde \y}^+_{-n}/n \right) b_{-j} ~\vr
\y^M, \y^L \rangle ~, ~~r \in -\bbn ~, \eqno(2.9a)$$ $$
\psi^{(-1)}_{-1,s} ~=~ \sum_{j=1}^{-s} S_{-s-j} \left( - {\tilde
\y}^-_{-n}/n \right) b_{-j} ~\vr \y^M, \y^L \rangle ~, ~~s \in -
\bbn ~. \eqno(2.9b)$$ Formulae (2.9a) for $r=-2$ and (2.9b) for
$s=-2$ were given in (1.19) as the states associated to the
operators (1.20).

{\bf 2.4.} ~~ The states considered above represent the extra
physical states for the {\bf (1,q) ~models} (or equivalently,
$(p,1)$ models), since in these cases all $(r,s)$ are on the
boundaries of the $(p=1) \times q$ rectangles of the conformal
lattice. (Of course, one has in addition as many states as these
taking into account the absolute cohomology.)

This follows from the more general observation that even for
arbitrary $(p,q)$ our formulae represent the extra physical
states for $(r,s)$ on the corners of the $p\times q$ rectangles
of the conformal lattice, i.e, for $r = kp$, $s = \ell q$,
$k,\ell \in\bbz$. Furthermore, our formulae are applicable also
for $(r,s)$ on the boundaries of the of the conformal lattice,
i.e, for $r = kp$, $s \neq \ell q$, or $s = \ell q$, $r \neq kp$,
$k, \ell \in\bbz$. We only have to change $\y^M \ra \y^L$, $u_\pm
\ra \g_\pm ((\y^L)^2 - (\x^L)^2) - 1$ in (2.2) - (2.5).

Our formulae are not directly applicable for $r,s$ in the
interiors of rectangles of the conformal lattice, i.e., for $r
\neq kp$, $s\neq \ell q$, $k,\ell \in\bbz$. The reason is that in
these cases the physical  states in $L (\x^M, \y^M) \otimes \cf
(\x^L, \y^L) \otimes \cf^G$, where $L(\x, \y)$ is the irreducible
highest weight module with highest weight $h = (\y^2 - \x^2)/2$,
can have arbitrary large ghost number (depending on $\x^L, \y^L$)
[5,11,14]).

\bigskip

{\bf Appendix. ~Action of Virasoro generators on Schur
polynomials}

Let us recall (cf., e.g., [18]) that the elementary Schur
polynomials $S_k$ are defined by the exponential generating
function: $$ exp \left( \sum_{k \in \bbn } t^k x_k \right) ~=~
\sum_{k \in \bbz} t^k S_k (x) ~, ~~~x ~=~ (x_1 , x_2 , \ldots) ,
\eqno(A.1)$$ which implies $$ S_k (x) ~=~ 0 ~, ~~k < 0 ~, ~~~~
S_0 (x) ~=~ 1 ~, \eqno(A.2a)$$ $$ S_k (x) ~=~ \sum_{k_1 + 2k_2 +
\ldots = k} ~{ x_1^{k_1} \over k_1 ! } ~{ x_2^{k_2} \over k_2 ! }
\ldots ~, ~~~k >0 ~. \eqno(A.2b)$$ For any partition $\l = \{
\l_1 \geq \l_2 \geq \ldots \}$ one associates a Schur polynomial:
$$S_\l (x) ~=~ S_{\l_1 , \l_2 , \ldots} (x) ~=~ {\rm det}
(S_{\l_j + k - j} (x))_{j,k} ~. \eqno(A.3)$$ For a given
partition $\l = \{ \l_1 \geq \l_2 \geq \ldots \geq \l_n \}$ with
$\l_n > 0$ we shall call $n$ the {\it length} of $\l$.

The Virasoro generators corresponding to the energy-momentum
tensor $T(z)$ are given in terms of the modes $\y_n$ as: $$ L_n
{}~=~ { 1 \over 2} \sum_{k \in \bbz} : \y_k \y_{n-k} : - (n+1) \x
\y_n ~, \eqno(A.4a)$$ where $\x$ is the background charge and $$[
\y_k ~, ~\y_n ] ~=~ k \d_{k, ~-n} ~. \eqno(A.4b)$$ The Fock
module $\cf (\y, \x)$ is determined by: $$ \y_n \vr \y \rangle
{}~=~ 0 ~, ~n > 0 ~, ~~~\y_0 \vr \y \rangle ~=~ (\y + \x) \vr \y
\rangle ~, \eqno(A.5a)$$ from which follows: $$ L_n \vr \y
\rangle ~=~ 0 ~, ~n > 0 ~, ~~~L_0 \vr \y \rangle ~=~ h \vr \y
\rangle ~, ~~ h ~=~ {1 \over 2 } (\y^2 - \x^2) ~. \eqno(A.5b)$$
{}From  (A.4) follows : $$[ L_n ~, ~\y_k ] ~=~ - k \y_{n+k} -
n(n+1) \x \d_{n, ~-k} ~. \eqno(A.6)$$ Denote for $k \neq 0$ ~$x_k
{}~=~ \g \y_{-k} /k $~, ~$\g = \g_\pm = \sqrt{2} \x_\pm$~, then we
have: $$[ L_n ~, ~x_k ] ~=~ (k-n) x_{k-n} + \g (\y - k\x)
\d_{n,k} ~, \eqno(A.7)$$ where $\y ~=~ \y_0 - \x$ anticipating
the action of $\y_0$ on $\vr \y \rangle$.

For the derivation of the singular vectors we need only the
action of $L_1$ and $L_2$ (since they generate all $L_n ~, ~n\geq
3$). ~Consider a function $f(x)$ of $x = (x_1, x_2, \ldots)$.
Then from (A.7) we obtain at once: $$[ L_1 , f(x)] ~=~
\sum_{j=1}^{\infty} j x_j { \pd f \over \pd x_{j+1} } + \sqrt{2}
\g ( \y - \x) { \pd f \over \pd x_1 } ~, \eqno(A.8)$$ and after
some manipulation: $$[ L_2 , f(x)] ~=~ \sum_{j=1}^{\infty} j x_j
{ \pd f \over \pd x_{j+2} } + \sqrt{2} \g ( \y - 2\x) { \pd f
\over \pd x_2 } + \g^2 { \pd^2 f \over \pd x_1^2 } +2\g^2 { \pd f
\over \pd x_1 } { \pd \over \pd x_1 } ~. \eqno(A.9)$$ The last
term in (A.9) indicates that $L_2$ is not a derivative operators
on functions of $x$.  We need the following properties of
elementary Schur polynomials [11] : $${ \pd S_k(x) \over \pd x_n
} ~=~ S_{n-k} ~, ~~~ \sum_{m = j+1}^k x_{m-j} S_{k-m} ~=~ (k-j)
S_{k-j} ~.\eqno(A.10)$$

Using these properties we obtain from (A.8), (A.9) the formulae
necessary for the derivation of the singular vectors: $$[ L_1 ,
S_k (x) ] ~=~ (k - 1 + \g ( \y - \x)) S_{k-1} (x) ~,
\eqno(A.11)$$ $$[ L_2 , S_k (x) ] ~=~ (k - 1 + \g ( \y - \x))
S_{k-2} (x) ~+~ \g^2 S_{k-1} { \pd \over \pd x_1 } ~.
\eqno(A.12)$$

As a direct application of (A.11) and (A.12) let us derive
(2.2a).  From (1.7) we have for $r=1$~: $$\g (\y^M - \x^M) ~=~
\g_+ (\y^M (1,s) - \x^M) ~=~ 1 - s ~, \eqno(A.13a)$$ and from
(A.11) and (A.12) we have : $$ [ L_n ~, ~S_k ] ~=~ (k-s) S_{k-n}
{}~~~ \Ra ~~~ L_n ~S_s (x) ~\vr \y^M \rangle ~=~ 0 ~, ~~n = 1,2 ~,
\eqno(A.13)$$ the last equality meaning that $S_s (x) ~\vr \y^M
\rangle$ is a singular vector.

For the action of $L_n$ on arbitrary Schur polynomials it is
convenient to use the following formulae which are derived easily
from (A.11) and (A.12): $$[ L_1 , S_{k_1 , \ldots , k_n} (x) ]
{}~=~ \sum_{j=1}^n ~(k_j - j + \g ( \y - \x)) ~S_{k_1, \ldots , k_j
- 1 , \ldots , k_n} (x) ~, \eqno(A.14)$$ $$\eqalign{ [ L_2 ,
S_{k_1 , \ldots , k_n} (x) ] ~&=~ \sum_{j=1}^n ~(k_j - j + \g (
\y - \x)) ~S_{k_1, \ldots , k_j - 2 , \ldots , k_n} (x) ~+ \cr
&+~ (\g^2 -1) \sum_{1 \leq i < j \leq n} S_{k_1, \ldots , k_i - 1
, \ldots , k_j - 1 , \ldots , k_n} (x) ~, \cr} \eqno(A.15)$$
where we have omitted the terms with derivatives which would act
if we apply $L_2$ to $S_{k_1 , \ldots , k_n} (x) ~f(x)$. Note
that application of (A.14),(A.15) involves Schur polynomials of
unordered partitions which are defined again by (A.3) and we have
to use: $$ S_{k_1, \ldots , k-1 , k , \ldots , k_n} ~=~ 0 ~,
{}~~~~S_{k_1, \ldots , k-2 , k , \ldots , k_n} ~=~ - S_{k_1, \ldots
, k-1 , k-1 , \ldots , k_n} ~. \eqno(A.16)$$

\bigskip

{\bf Acknowledgments}

The authors would like to thank E. Gava, K.S. Narain and M.H.
Sarmadi for discussions. They would like also to thank Professor
Abdus Salam for hospitality at the ICTP.

\bigskip

\baselineskip=12pt

{\bf References}

\item{[1]} A.M. Polyakov, Phys. Lett. {\bf 103B} (1981) 207; Mod.
Phys. Lett. {\bf A2} (1987) 893.

\item{[2]} V.G. Knizhnik, A.M. Polyakov and A.B. Zamolodchikov,
Mod. Phys. Lett. {\bf A3} (1988) 819.

\item{[3]} J. Distler and H. Kawai, Nucl. Phys. {\bf B321} (1989)
509.

\item{[4]} F. David, Mod. Phys. Lett. {\bf A3} (1988) 1651.

\item{[5]} B.H. Lian and G.J. Zuckerman, Phys. Lett. {\bf 254B}
(1991) 417.

\item{[6]} B.H. Lian and G.J. Zuckerman, Phys. Lett. {\bf 266B}
(1991) 21.

\item{[7]} A.M. Polyakov, Mod. Phys. Lett. {\bf A6} (1991) 635.

\item{[8]} S. Mukherji, S. Mukhi and A. Sen, Phys. Lett. {\bf
266B} (1991) 337.

\item{[9]} C. Imbimbo, S. Mahapatra and S. Mukhi, preprint INFN
Genova \& Tata Institute, GEF-TH 8/91 \& TIFR/TH/91-27 (May
1991).

\item{[10]} K. Itoh, preprint CTP-TAMU-42/91 (June 1991).

\item{[11]} P. Bouwknegt, J. McCarthy and K. Pilch, preprints
CERN-TH.6162/91 (July 1991) \& CERN-TH.6196/91 (August 1991).

\item{[12]} E. Witten, preprint IASSNS-HEP-91/51 (August 1991).

\item{[13]} I.R. Klebanov and A.M. Polyakov, Mod. Phys. Lett.
{\bf A6} (1991) 3273.

\item{[14]} P. Bouwknegt, J. McCarthy and K. Pilch, preprint
CERN-TH.6279/91 (October 1991).

\item{[15]} M. Goulian and M. Li, Phys. Rev. Lett. {\bf 66}
(1991) 2051; ~P. Di Francesco and D. Kutasov, Phys. Lett. {\bf
261B} (1991) 385; ~Y. Kitazawa, Phys. Lett. {\bf 265B} (1991)
262; ~N. Sakai and Y. Tanii, Prog. Theor. Phys. {\bf 86} (1991)
547;  ~L. Alvarez-Gaum\'{e}, J.L.F. Barb\'{o}n and C. G\'{o}mez,
Nucl. Phys. {\bf B368} (1992) 57; ~L. Alvarez-Gaum\'{e} and C.
G\'{o}mez,  preprint CERN-TH.6175/91 (July 1991); ~Vl.S.
Dotsenko, preprint Paris VI PAR-LPTHE 91-18 (October 1991); ~M.
Li, preprint UCSBTH-91-47 (October 1991).

\item{[16]} D. Kutasov, E. Martinec and N. Seiberg, preprint
PUPT-1293 (November 1991).

\item{[17]} Y. Cai and G. Siopsis, Mod. Phys. Lett. {\bf A6}
(1991) 1261.

\item{[18]} J. Goldstone, unpublished; ~G. Segal, Comm. Math.
Phys. {\bf 80} (1981) 301.

\item{[19]} I. Bakas, Phys. Lett. {\bf 228B} (1989) 57; ~C.N.
Pope, L.J. Romans and X. Shen, Phys. Lett. {\bf 236B}  (1990)
173.

\item{[20]} B.L. Feigin and D.B. Fuchs, Representations of the
Virasoro algebra, Seminar on supermanifolds No. 5, Ed. D. Leites
(1985).

\item{[21]} V.G. Kac, Lecture Notes in Physics, Vol. 94 (1979)
pp.441--445.

\item{[22]} A.A. Belavin, A.M. Polyakov and A.B. Zamolodchikov,
Nucl. Phys. {\bf 241} (9184) 333.

\item{[23]} B.L. Feigin and D.B. Fuchs, Funkts. Anal. Prilozh.
{\bf 17} (3) 91-92 (1983).

\item{[24]} V.K. Dobrev, in: Proc. XIII Int. Conf. Diff.-Geom.
Meth. Theor. Phys., (Shumen, 1984), Eds. H.D. Doebner and
T.D.Palev (World Sci., Singapore, 1986) p. 348.

\vfil\eject\end